\documentclass[aps,prd,showpacs,twocolumn,nofootinbib,superscriptaddress]{revtex4}
\usepackage{cancel}
\usepackage{amssymb,amsfonts,amsmath}
\usepackage{graphicx,epsfig,pdfpages}
\newcommand{\bea}{\begin{eqnarray}}
\newcommand{\eea}{\end{eqnarray}}

\begin{document}
\hspace*{5.85in} \hbox{TIFR-TH/15-47} 
\title{Diphoton resonance at 750 GeV in the broken MRSSM}

\author{Sabyasachi Chakraborty}
\email{sabya@theory.tifr.res.in}
\author{Amit Chakraborty}
\email{amit@theory.tifr.res.in}
\author{Sreerup Raychaudhuri}
\email{sreerup@theory.tifr.res.in}
\affiliation{Department of Theoretical Physics, Tata Institute of Fundamental 
Research, \\ 1, Homi Bhabha Road, Mumbai 400005, India.}

\begin{abstract}
Non-observation of superpartners of the Standard Model particles at the 
early runs of the LHC provide strong motivation for an $R$-symmetric 
minimal supersymmetric Standard Model, or MRSSM. This model also comes with 
a pair of extra scalars which couple only to superpartners at the tree 
level. We demonstrate that in the limit when the $U(1)_R$ symmetry is 
broken, one of these scalars develops all the properties necessary to 
explain the 750~GeV diphoton resonance recently observed at the LHC, as 
well as the non-observation of associated signals in other channels. Some 
confirmatory tests in the upcoming LHC runs are proposed.
\end{abstract}


\maketitle
\section{Introduction}
\label{intro}

In many ways, supersymmetric models remain the best option for new physics 
beyond the Standard Model (SM) of electroweak and strong interactions. The 
discovery of a light, probably elementary, scalar in 2012 \cite{Higgs2012} 
has made this motivation, if anything, stronger than ever. However, this 
must be coupled with the somewhat disappointing fact that the early runs of 
the Large Hadron Collider (LHC) at CERN, Geneva, have not found any of the 
promised signals for supersymmetry (SUSY)\cite{susycms,susyatlas}. 
Moreover, the decay modes of the 125~GeV scalar found in 2012 appear 
increasingly to resemble those of the SM Higgs boson 
\cite{Signalstrengths}. Though all this does not invalidate the idea of 
SUSY {\it per se}, it has made it increasingly difficult to fit the 
observed results with popular models of SUSY, such as the so-called minimal 
supersymmetric SM, or MSSM.

To add to this tension, we have the recent announcement that both the ATLAS 
and CMS Collaborations seem to have observed~\cite{ATLAS,CMS} an excess of 
diphoton events in the 13~TeV run, which commenced earlier this year. The 
excess events appear to arise from a resonant production of an intermediate 
scalar particle of mass around $750$~GeV and a width which is best-fitted 
as 45~GeV. At the same time, both the experimental collaborations have 
announced that searches for deviations from the SM prediction in all other 
channels have produced null results. Their principal results on the 
diphoton excess are summarised below.
\begin{itemize}
\item The ATLAS Collaboration has observed \cite{ATLAS} an excess of 14 
events, with a peak at 750~GeV and a best-fit width of 45~GeV, in 
$3.2$~fb$^{-1}$ of data at $\sqrt{s} = 13$~TeV. The local significance of 
this excess is $3.9\sigma$, but it reduces to about $2.6\sigma$ if the 
look-elsewhere effect is included. Taking into account the experimental 
acceptance value of about 0.4, this corresponds to an excess signal of $10 
\pm 3$~fb.
\item The CMS Collaboration has observed \cite{CMS} an excess of 10 events, 
with a peak at 760~GeV, in $2.6$~fb$^{-1}$ of data at $\sqrt{s} = 13$~TeV. 
The local significance of this excess is $2.3\sigma$, but it reduces to 
about $2.0\sigma$ if the width is assumed to be around 45~GeV. Taking into 
account the experimental acceptance value of about 40\%, this corresponds 
to an excess signal of $6\pm 3$~fb.
\end{itemize}
While there is a strong probability that this excess is only a statistical 
fluctuation in the data, there is always the exciting possibility that this 
may be the first observed manifestation of new physics at the LHC -- or, 
for that matter, any other collider experiment. Undoubtedly, this 
announcement has stirred the theoretical mind, for several new physics 
interpretations of this excess have already appeared in the literature. For 
example, models with vector-like fermions and extended scalar 
sectors~\cite{scalar1,scalar2,VF,Megias:2015ory, 
Liao:2015tow,Bi:2015uqd,Berthier:2015vbb}, SUSY~\cite{scalar1,susy750}, 
extra dimensions~\cite{Megias:2015ory,radion}, axions and composite 
scalars~\cite{axion,Franceschini:2015kwy}, vector 
resonances~\cite{Chala:2015cev}, leptoquarks~\cite{Murphy:2015kag}, dark 
matter candidates~\cite{Bi:2015uqd,DM}, minimal gauge extensions of the SM 
and MSSM~\cite{MGE} have been studied. Some have proposed model-independent 
tests of the signal \cite{Franceschini:2015kwy,MI}, and others have 
constructed scenarios in which the presence of a diphoton excess and the 
absence of any other signals arises in a natural way~\cite{others}. In 
addition, electroweak vacuum stability and inflation in the presence of 
this new resonance has been analysed in Refs.\cite{Dhuria}.  However, it is 
probably a fair statement to say that an explanation of the current results 
is rather difficult to obtain in any of the popular `minimal' models which 
have hitherto been the mainstay of phenomenological studies of physics 
beyond the SM. Quite naturally, therefore, many of the proposed scenarios 
invoke exotic options, which are barely permitted by the experimental data 
and do not conform to the choices commonly seen earlier in the 
literature~\cite{Jaeckel:2012yz}. It is interesting, therefore, to ask, if 
there can be found a well-motivated model, where a specific scenario in the 
parameter choices could explain the observed facts in this regard.

In this article, we consider the minimal $R$-symmetric supersymmetric SM, 
or MRSSM~\cite{Rsym}, which -- apart from the usual virtues of a SUSY model 
-- can explain in a natural way, the non-observance of SUSY-specific 
signals at the LHC in the present and previous runs. These models generally 
contain Dirac gauginos in their spectra as opposed to Majorana gauginos in 
MSSM. The presence of a Dirac gluino reduces the production cross-section 
for squarks considerably, explaining their absence in LHC data. Multiple 
versions of models with Dirac gauginos can be found in the literature on 
SUSY~\cite{Fayet:1976et}. In these, flavor and $CP$-violation constraints 
are suppressed~\cite{Kribs:2007ac} and issues pertaining to neutrino mass 
generation and dark matter can also be addressed 
\cite{Chakraborty:2013gea,Chakraborty:2014tma}. To cut a long story short, 
once we have Dirac gauginos in a $R$-symmetric model, it becomes necessary 
to incorporate two additional $SU(2)$-doublet chiral superfields $\widehat 
R_u$ and $\widehat R_d$ carrying nonzero $R$-charges. To avoid spontaneous 
$R$-breaking and the emergence of $R$-axions, the scalar components of 
$\widehat R_u$ and $\widehat R_d$ do not receive any nonzero vacuum 
expectation value (vev). Hence they are dubbed `inert' doublets. It is one 
of these `inert' scalars which we propose as a candidate for the 750 GeV 
resonance.

The plan of this paper is as follows. In Section~II, we describe the 
$R$-symmetric version of the MSSM, illustrating the role of the `inert' 
doublets mentioned above. We then go on, in Section~III, to explain how 
$R$-symmetry requires to be broken in order to obtain a left-right mixing 
in the top-squark sector, which is vital to get a diphoton signal. 
Section~IV is devoted to an explanation of how the diphoton excess arises 
in this model. In Section~V, we summarise our results and mention some 
tests which may falsify this scenario in future runs of the LHC.

\section{MRSSM -- the framework}
\label{framework}

In $R$-symmetric models, one adds to the symmetries of the SM an extra 
$U(1)_R$ global symmetry, under which the superpartner fields transform, 
but the SM fields do not. This $R$-symmetry prohibits Majorana gaugino 
masses, trilinear scalar couplings and forces us to set the Higgsino mass 
parameter $\mu = 0$. Hence, the gauginos need to be Dirac fermions, to 
construct which one needs to introduce additional superfields, such as a 
color and $SU(2)_L$ singlet $\widehat S$, a colorless $SU(2)_L$ triplet 
$\widehat T$ and another $SU(2)_L$ triplet $\widehat{\mathcal O}$ which 
transforms as an octet under $SU(3)_c$. An immediate consequence of this is 
that squarks coupling to quarks and a Dirac gluino have much lower 
production cross-sections at the LHC than they would have had in the usual 
case of a Majorana gluino. This significantly weakens the rather tight 
constraints on light squarks which have already been obtained at the LHC. 
To ensure, however, that the lighter chargino $\widetilde{\chi}_1^\pm$ does 
not become massless, we require to generate a $\mu$ term by adding two new 
superfields $\widehat R_u$ and $\widehat R_d$ carrying non-zero 
$R$-charges. The SM gauge quantum numbers and $U(1)_R$ charges of all the 
chiral superfields in the model are shown in Table~\ref{table:1}.
\begin{table}[htb]
\centering
 \begin{tabular}{ccc} 
 \hline
 Superfields & SM rep & $U(1)_R$ \\ [0.5ex] 
 \hline\hline
 $\widehat Q_i$ & $(3,2,\frac{1}{3})$ & 1 \\ 
 $\widehat U_i^c$  & $(\bar 3,1,-\frac{4}{3})$ & 1 \\
 $\widehat D_i^c$ & $(\bar 3,1,\frac{2}{3})$ & 1 \\
 \hline
 $\widehat L_i$ & $(1,2,-1)$ & 1 \\
 $\widehat E_i^c$ & $(1,1,2)$ & 1 \\
 \hline
 $\widehat H_u$ & $(1,2,1)$ & 0 \\
 $\widehat H_d$ & $(1,2,-1)$ & 0 \\
 \hline
 $\widehat R_u$ & $(1,2,-1)$ & 2 \\
 $\widehat R_d$ & $(1,2,1)$ & 2 \\
 \hline
 $\widehat S$ & $(1,1,0)$ & 0 \\ 
 $\widehat T$  & $(1,3,0)$ & 0 \\
 $\widehat{\mathcal O}$ & $(8,1,0)$ & 0 \\
 \hline
\end{tabular}
\caption{The chiral superfields in the MRSSM, showing their gauge 
quantum numbers under the SM gauge group $SU(3)_c \times SU(2)_L \times 
U(1)_Y$ as well as their $U(1)_R$ charge assignments.}
\label{table:1}
\end{table}
It is important to note that the scalars $R_u$ and $R_d$ have the same 
$R$-charge as the superfields $\widehat{R}_u$ and $\widehat{R}_d$ whereas 
the $R$-charges of the fermions $\widetilde{R}_u$ and $\widetilde{R}_d$ are 
less by one unit. In addition, to have an invariant action, the 
superpotential has to have $R$-charge of two units. This superpotential can 
now be written as
\begin{eqnarray}
W&=& 
\mu_d \widehat R_d\widehat H_d +\mu_u\widehat R_u\widehat H_u \nonumber \\
&+&\Lambda_d\widehat R_d \widehat T\widehat H_d
+\Lambda_u \widehat R_u\widehat T\widehat H_u
+\lambda_d\widehat S\widehat R_d\widehat H_d
+\lambda_u\widehat S\widehat R_u\widehat H_u \nonumber \\
&+&Y_d\widehat Q_i \widehat H_d\widehat D_i^c
+Y_e\widehat E_i^c\widehat L_i\widehat H_d
+Y_u\widehat U_i^c \widehat Q_i\widehat H_u.
\end{eqnarray}
where the $\mu$'s, $\Lambda$'s, $\lambda$'s and $Y$'s are constants.

This model now has an extended Higgs sector as well as an extended 
fermionic sector. If the vev's of the neutral scalar component of the 
superfields $\widehat S$ and $\widehat T$ are small, the corresponding 
scalars can be integrated out from the theory and one is left only with a 
SM-like $H$ doublet and the `inert' doublets $R_u$ and $R_d$. The neutral 
part of the potential, consisting of $F$ terms, $D$ terms and explicit soft 
SUSY-breaking terms, can then be written as
\begin{eqnarray}
V_{\textrm{neut}} 
&=& (m^2_{H_d}+\mu_d^2)|H_d^0|^2+(m^2_{H_u}+\mu_u^2)|H_u^0|^2\nonumber \\
&+&\frac{1}{8}(g^2+g^{\prime 2})(|H_d^0|^2-|H_u^0|^2-|R_d^0|^2+|R_u^0|^2) 
\nonumber \\ 
&-&(m^2_{R_u}+\mu_u^2)|R_u^0|^2+(m^2_{R_d}+\mu_d^2)|R_d^0|^2\nonumber \\
&+&(B\mu H_d^0 H_u^0+\textrm{h.c.})+|\lambda_u R_u^0 H_u^0
-\lambda_d R_d^0 H_d^0|^2 \nonumber \\
&+& \Big|\frac{\Lambda_d}{\sqrt 2}R_d^0 H_d^0
+\frac{\Lambda_u}{\sqrt 2}R_u^0 H_u^0\Big|^2 \ .
\label{scalarpot}
\end{eqnarray}
This potential can now be minimised to find the scalar eigenstates of the 
model. It is important to note that after electroweak symmetry breaking, 
the $R_u^0$ and $R_d^0$ scalars mix with each other, but not with the $H^0$ 
state. Moreover, the $R$-charge assignments of these $R$-scalars restricts 
their trilinear couplings only to ($a$) sfermions and chargino/neutralino 
combinations, e.g.  $R\widetilde{\ell}\widetilde{\ell}$, 
$R\widetilde{q}\widetilde{q}$, $R\widetilde{\chi}\widetilde{\chi}$, and 
($b$) paired-$R$ scalars to SM bosons, i.e. $RRH$ and $RRV$, where $V = 
W^\pm,Z^0$. $R$-scalar couplings to pairs of any SM particle vanish. 
$R$-scalar couplings to sfermions, which play a major role in our work, are
\begin{eqnarray}
{\mathcal L}_{R\tilde{f}\tilde{f}^*} 
& = & -\mu_u Y_u R_u^0 \widetilde u_R\widetilde u_L^* 
-\mu_d Y_d R_d^0 \widetilde d_R\widetilde d_L^* \nonumber \\
&&  -\mu_d Y_e R_d^0 \widetilde e_R \widetilde e_L^*.
\end{eqnarray}
where $Y_u$, $Y_d$ and $Y_e$ are Yukawa couplings of the SM and a sum over 
generations is implicit. For third generation quarks we have $Y_t \gg Y_b$ 
and therefore we will mostly confine ourselves to the $R_u^0$ scalar. It is 
important to note that the $R_u^0$ scalar couples only to $\widetilde 
q_L$--$\widetilde q_R^*$ pairs, and not to $\widetilde q_L$--$\widetilde 
q_L^*$ or $\widetilde q_R$--$\widetilde q_R^*$ pairs. As a result, in the 
$R$-conserving scenario, the $R_u^0$ scalar cannot couple to photon pairs 
through top-squark loops as there is no mixing between the $\widetilde t_L$ 
and $\widetilde t_R$ states. It is clear, therefore, that a diphoton signal 
from decay of a resonant $R_u^0$ requires us to break $R$-symmetry.

\section{$R$-symmetry breaking}
\label{R-breaking}

In addition to the phenomenological need mentioned in the previous section, 
there exist strong motivations for the breaking of $R$-symmetry from 
cosmological considerations \cite{SPMartin}. Assuming, therefore, that the 
$R$-symmetry breaks spontaneously in the hidden sector (like supersymmetry) 
the $R$-breaking information must be communicated to the visible sector by 
some mechanism such as gravity mediation, anomaly mediation, etc. For our 
purposes, we do not require to consider a particular breaking mechanism, 
but it suffices to parametrise the $R$-breaking information in terms of a 
set of trilinear scalar couplings (which also break SUSY). In fact, for an 
$R_u^0$-scalar decaying to two photons via top-squark loops, the only 
relevant $R$-breaking term in the Lagrangian is given as
\begin{eqnarray}
\mathcal L_{\cancel{R}} = A_t H_u \widetilde Q_3\widetilde U_3^c,
\end{eqnarray}
where $A_t$ is the trilinear scalar coupling. After electroweak 
symmetry-breaking, this term generates a mixing between the left- and the 
right-chiral top-squarks. The mass-squared matrix for the top-squarks takes 
the form
\begin{equation}
{\cal M}^2_{\widetilde{t}} = \left[ \begin{array}{cc} 
({\cal M}^2_{\widetilde{t}})_{11}  & ({\cal M}^2_{\widetilde{t}})_{12} \\
({\cal M}^2_{\widetilde{t}})_{21} & ({\cal M}^2_{\widetilde{t}})_{22} 
\end{array} \right]
\end{equation}
where
\begin{eqnarray}
({\cal M}^2_{\widetilde{t}})_{11} &=& 
\frac{1}{8} \Big(g^2 + \frac{g^{\prime 2}}{3}\Big)(v_d^2-v_u^2)+
m^2_{\widetilde t_L}+\frac{1}{2}Y_t^2 v_u^2, \nonumber \\
({\cal M}^2_{\widetilde{t}})_{12} &=& ({\cal M}^2_{\widetilde{t}})_{21} 
= A_t\, v_u, \nonumber \\
({\cal M}^2_{\widetilde{t}})_{22} &=& \frac{g^{\prime 2}}{6}(v_d^2-v_u^2)+
m^2_{\widetilde t_R}+\frac{1}{2}Y_t^2 v_u^2.
\end{eqnarray}
in terms of the vev's $v_u$ and $v_d$ of the two Higgs doublets $H_u$ and 
$H_d$ respectively. The mixing angle $\theta_{\widetilde{t}}$ is now given 
by
\begin{eqnarray}
\tan 2\theta_{\widetilde{t}} &=& 
\frac{2 v_u A_t}{\frac{1}{8}(v_d^2-v_u^2)(g^2-g^{\prime 2})+
(m^2_{\widetilde t_L}-m^2_{\widetilde t_R})}.
\end{eqnarray}
It is amusing to note that one can generate maximal mixing even without 
taking the $R$-breaking parameter $A_t$ to be unnaturally large, for the 
same effect can be obtained by setting $v_u \approx v_d$ and 
$m^2_{\widetilde t_L} \simeq m^2_{\widetilde t_R}$.

This mixing between $\widetilde t_L$ and $\widetilde t_R$ is crucial for 
our analysis, since it permits the $R_u^0$-scalar to decay to diphotons and 
to be produced through gluon fusion by top-squark loops --- which would not 
be possible otherwise, as explained in the previous section.

\section{Fitting the Diphoton signal}
\label{twophoton}

\begin{figure}
\includegraphics[scale=0.35]{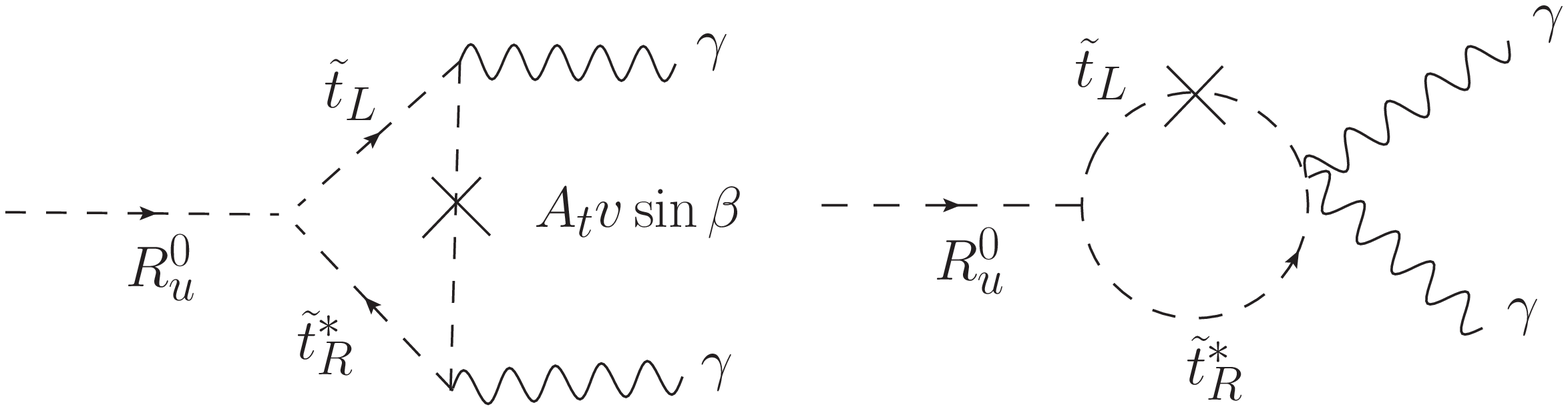}
\caption{\small Top-squark loops contributing to $R_u^0 \to \gamma\gamma$. 
Similar diagrams mediate gluon-gluon fusion $gg \to R_u^0$, where the 
photon lines are replaced by gluon lines.} \label{fig:1}
\end{figure}

The decay of the $R_u^0$-scalar to a $\gamma\gamma$ pair is mediated at the 
one-loop level dominantly by the diagrams shown in Figure~\ref{fig:1} 
(including a crossed diagram). Similar diagrams exist for its decay to a 
$gg$ pair.
Evaluation of these diagrams leads to the partial widths
\begin{eqnarray}
\Gamma(R\rightarrow\gamma\gamma) &\simeq& \frac{\alpha^2 N_c^2 Q_{\widetilde t}^4}
{1024\pi^3}\frac{M_R^3\mu_{\rm eff}^2}{M^4_{\widetilde t}}\, |F(\tau)|^2 
\nonumber \\
\Gamma(R\rightarrow gg) &\simeq& \frac{\alpha_s^2}{512\pi^3}
\frac{M_R^3\mu_{\rm eff}^2}{M_{\widetilde t}^4}\,|F(\tau)|^2
\label{eqn:widths}
\end{eqnarray}
where $\alpha$ and $\alpha_s$ are the electromagnetic and strong coupling 
constants, $N_c$ is the colour factor, $Q_{\widetilde t}$ = 2/3 is the 
fractional charge of the top-squark and $M_R$ is the mass of the 
$R_u^0$-scalar. In the above formulae, $\mu_{\rm eff}$ is an effective 
coupling defined as
\begin{eqnarray}
\mu_{\rm eff} = \frac{\mu_u Y_t}{4}\sin^2 2\theta_{\widetilde{t}},
\end{eqnarray}
and $F(\tau)$ is the loop integral function
\begin{equation}
F(\tau) = \left(\tau \sin^{-1} \frac{1}{\sqrt{\tau}} \right)^2 - \tau
\end{equation}
where $\tau = 4M^2_{\widetilde t}/M_R^2$. Here, $M_{\widetilde t}$ is the 
mass of the lighter eigenstate of the top squark. This particular form of 
$F(\tau)$ arises only in the case $2M_{\widetilde t} > M_R$, which is 
assumed by us to ensure that the $R_u^0$ does not decay at the tree level 
to a pair of top-squarks.

We are now in a position to compare the predictions of this model with the 
experimental results quoted in the Introduction. It is necessary to point 
out, at this stage itself, that we assume that all tree-level decays of the 
$R_u^0$ scalar are kinematically disallowed. The spectrum of superparticles 
can be chosen to satisfy this without conflicting with any known 
theoretical or experimental requirements.

It is most convenient to treat the two widths $\Gamma_{\gamma\gamma} = 
\Gamma(R_u^0 \to \gamma\gamma)$ and $\Gamma_{gg} = \Gamma(R_u^0 \to gg)$ as 
correlated variables, and study the plane formed by plotting them against 
each other. The production cross-section for the $R_u^0$ scalar will be 
given in terms of $\Gamma_{gg}$ by \begin{equation} \sigma_R = 
\frac{\pi^2}{8} \frac{\Gamma_{gg} C_{gg} K_{gg}}{sM_R} \end{equation} where 
$C_{gg}$ is the gluon density function given by \begin{equation} C_{gg} = 
\int_\delta^1 \frac{dx}{x} \, 
f_{g/p}(x)\,f_{g/p}\left(\frac{\delta}{x}\right) \end{equation} with 
$\delta = M_R^2/s$, where $\sqrt{s} = 13$~TeV, the machine energy. The 
functions $f_{g/p}(x)$ are, of course, the gluon parton-density functions. 
$K_{gg}$ is a QCD correction factor which we take to be approximately 1.5 
\cite{Kfactor}. In fact, using the CTEQ-6 \cite{Pumplin:2002vw} set of 
structure functions, we evaluate $C_{gg} \approx 2914$, from which, it 
follows that the production cross-section is \begin{equation} \sigma_R 
\approx 12.4~{\rm nb} \times \frac{\Gamma_{gg}}{M_R} \end{equation} 
\begin{figure} 
\includegraphics[scale=0.7]{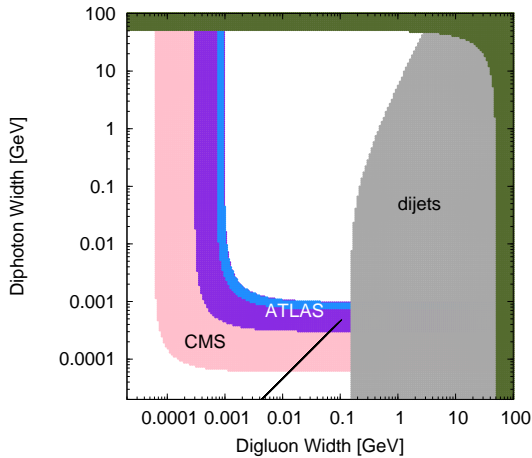} 
\caption{\small Illustrating 
constraints on the $\Gamma_{\gamma\gamma}$--$\Gamma_{gg}$ plane from the 
experimental data as well as the predictions of the MRSSM model.} 
\label{fig:2} 
\end{figure}  
We now have the cross-section for 
\begin{itemize} 
\item diphotons, given by 
\begin{equation} 
\sigma_{\gamma\gamma} = \sigma_R \frac{\Gamma_{\gamma\gamma}}{\Gamma_R} 
\end{equation} 
where $\Gamma_R = \Gamma_{gg} + \Gamma_{\gamma\gamma}$ is the total width 
of of the $R_u^0$ resonance, assuming that no other decay modes are 
available to the $R_u^0$ scalar --- which will be the case if 
$2M_{\widetilde t} > M_R$, as assumed. \item dijets, given by
\begin{equation} 
\sigma_{gg} = \sigma_R \frac{\Gamma_{gg}}{\Gamma_R} 
\end{equation} 
As the $R_u^0$ scalar has no coupling with quarks and it is lighter than 
all the squark pairs, we can safely assume that the decay of a $R_u^0$ to 
dijets is completely dominated by the $gg$ mode.
\end{itemize}

Our analysis is then based on the following constraints.
\begin{enumerate}
\item
The total width $\Gamma_R$ of the $R_u^0$ scalar should satisfy
\begin{equation}
\Gamma_R < 50~{\rm GeV}
\label{eqn:cwid}
\end{equation}
Since the best-fit width is about 45~GeV, the value 50~GeV chosen above
seems to provide a reasonable leeway for errors.
\item
The dijet cross-section observed at the LHC in the 13~TeV run is consistent 
with the SM prediction of about $12.5 \pm 1.2$~pb \cite{ATLASdijet}. Thus, 
we must demand that the dijet excess arising from decay of the $R_u^0$ 
satisfies
\begin{equation}
\sigma_{gg} < 2.5~{\rm pb}
\label{eqn:cdij}
\end{equation}
assuming agreement with the SM at the 95\% confidence level. 
\item
The diphoton excess must be consistent with the observed values as 
presented by the ATLAS and CMS Collaborations (see Introduction). If we 
consider the 95\% confidence level, the ATLAS results require
\begin{equation}
4~{\rm fb} < \sigma_{\gamma\gamma} < 16~{\rm fb}
\label{eqn:catl}
\end{equation}
and the CMS results may be taken to require
\begin{equation}
1~{\rm fb} < \sigma_{\gamma\gamma} < 12~{\rm fb}
\label{eqn:ccms}
\end{equation}
\end{enumerate}

Combining all these constraints, we display our results in 
Figure~\ref{fig:2}, which shows the $\Gamma_{\gamma\gamma}$--$\Gamma_{gg}$ 
plane for a wide range of values from $10^{-5}$ to $10^2$. The dark-green 
shaded strip along the top and right of this panel represents the range 
ruled out by the total width constraint in Eqn.~(\ref{eqn:cwid}). The 
larger region on the right side of the panel, shaded grey, represents the 
dijet constraint in Eqn.~(\ref{eqn:cdij}), i.e. all points in the region 
would lead to an observable dijet signal at the 13~TeV run, which is not 
the case. The L-shaped regions depict the regions {\it allowed} by the 
ATLAS (blue) and CMS (pink) observations, with the overlap region appearing 
purple. Obviously, the two ends of each strip indicate either a large 
$\Gamma_{gg}$ with a small $\gamma\gamma$ branching ratio, or a small 
$\Gamma_{gg}$ (i.e. a small production cross-section) but a $\gamma\gamma$ 
branching ratio almost unity.

The oblique black line close to the lower end of Figure~\ref{fig:2} 
represents the predictions of the MRSSM model, as we vary $\mu_{\rm eff}$ 
up to a value of 2.5~TeV (which is well within the perturbative limit) and 
the (lighter) top-squark mass from $M_R/2$ to about 1~TeV. It is 
immediately clear that the predictions are nicely consistent with both the 
ATLAS and CMS observations, as the line passes clearly through both the 
allowed strips. We may claim, therefore, to have a neat explanation of the 
observed diphoton excess (and the absence of other signals) in the MRSSM, 
without having had to extend the field content specifically for this 
purpose.

We note, however, that this MRSSM solution leads to the prediction of a 
somewhat low width of 100~MeV or less for the $R_u^0$ resonance. This, 
while definitely larger than the Higgs boson width in the SM (4 MeV), is 
still small compared to the widths of the $W$ and $Z$ bosons. We can 
attribute the long life of the $R_u^0$ to the fact that it can only decay 
through one-loop diagrams. After all, it is an `inert' scalar! A small 
decay width is not a problem for the model at this stage of 
experimentation, since the kind of low statistics available at the moment 
leads to very poor estimations of the decay width. It is also important to 
note that larger widths of 200 MeV or more are incompatible with the 
non-observation of a dijet excess --- this is a generic feature of models 
having a scalar decaying exclusively to $\gamma\gamma$ and $gg$ modes.
 
\section{Critical Summary}
\label{conclusion}

In this article, therefore, we have shown that among the various possible 
explanations of the diphoton excess observed at the LHC, there exists the 
possibility of a SUSY solution which invokes an extra symmetry -- the 
$R$-symmetry -- but does not require us to postulate new fields 
specifically to explain the effect. Apart from introducing a pair of new 
scalars and some superfields to convert the gauginos from Majorana to Dirac 
fermions, this model retains the MSSM field content. However, we also 
obtain a good explanation of the failure of LHC to discover SUSY signals 
till date. We also require the $R$-symmetry to be broken by a solitary 
scalar trilinear operator, for otherwise the `inert' scalars could not be 
produced at all in hadron-hadron collisions.

An obvious question to be asked before concluding this analysis is whether 
there are any confirmatory tests which could be used to verify if the ideas 
presented here are indeed correct. This can be answered quite easily in the 
affirmative. We argue as follows. The straight line shown in 
Figure~\ref{fig:2} enters the allowed region only if the (lighter) 
top-squark has a mass in the range of a few hundred GeV, which would bring 
it very much within the kinematic range accessible for discovery at the LHC 
Run-2. Moreover, the neutral scalars $R_u^0$ and $R_d^0$ will be 
accompanied by their charged counterparts $R_u^\pm$ and $R_d^\pm$, and one 
could expect the mass ranges not to be very different. Charged scalars, of 
course, are easy to detect, and if they lie within the kinematic range of 
LHC (as we have good reason to suspect), it cannot be long before they will 
be discovered. Thus, we have a couple of very clear ways in which the model 
in question can be falsified. The truth will only be known when more data 
are acquired and analysed, but, for the moment, we may rest satisfied that 
the MRSSM has enough pleasing features to be taken very seriously as an 
explanation of the recent LHC enigma.

\acknowledgements
The authors would like to thank Disha Bhatia and Tuhin S. Roy for 
discussions. The work of SR is partly funded by the Board of Research in 
Nuclear Studies, Government of India, under project no. 2013/37C/37/BRNS.


\end{document}